\documentclass[12pt]{article}
\pdfoutput=1 

\usepackage{graphicx}
\usepackage{url}
\usepackage{hyperref}
\usepackage{amsmath}
\usepackage{amssymb}
\usepackage{natbib}
\usepackage{xcolor}

\usepackage[hmargin=0.75in,vmargin=1in]{geometry}
\title{Radial Outflow Explains the Rotation Curves of Disk Galaxies}

\date{September 2023}

\author{Earl Schulz}

\author{Earl Schulz\thanks{%
(Independent Researcher; Granby, Ct, USA; email:{earlschulz@gmail.com})}}

\def\araa{Annu. Rev. Astron. Astrophys.} % Annual Review of Astron and Astrophys
\def\aj{Astron. J.} % Astronomical Journal
\def\apj{Astrophys. J.} % Astrophysical Journal
\def\apjl{Astrophys. J. Lett.} % Astrophysical Journal, Letters
\def\aap{Astron. Astrophys.} % Astronomy and Astrophysics
\def\aapr{Astron. Astrophys. Rev.} % Astronomy and Astrophysics Reviews
\def\bain{Bull. Astron. Inst. Netherlands}
\def\mnras{Mon. Not. R. Astron. Soc.} % Monthly Notices of the RAS
\def\na{New Astron.} % New Astronomy
\def\prd{Phys. Rev. D} % Physical Review D
\def\pre{Phys. Rev. E} % Physical Review E
\def\pasa{Publ. Astron. Soc. Aust.} % Publications of the Astron. Soc. of Australia
\def\pasp{Publ. Astron. Soc. Pac.} % Publications of the Astron. Soc. of the Pacific
\newcommand{\aeff}    {\ensuremath{a_\textrm{eff}}}
\newcommand{\aMil}    {\ensuremath{a_0}}
\newcommand{\aN}      {\ensuremath{a_\textrm{N}}}
\newcommand{\app}     {\ensuremath{\sim}}
\newcommand{\CAF}     {\ensuremath{C_\textrm{AF}}}

\newcommand{\dex}[1]  {\ensuremath{\times\textrm{10}^{#1}}}

\newcommand{\Etot}    {\ensuremath{E_\textrm{tot}}}
\newcommand{\FMD}     {\ensuremath{ \textrm{FMD} }}

\newcommand{\Ha}      {\textrm{H}\ensuremath{ {\alpha}}}
\newcommand{\HI}      {\textrm{H{\sc i}}}
\newcommand{\LCDM}    {\ensuremath{\Lambda\textrm{CDM}}}
\newcommand{\Lsun}    {\ensuremath{ \textrm{L}_{\odot}}}
\newcommand{\mas}     {\ensuremath{\textrm{ mag arcsec}^{-2}}}

\newcommand{\mm}      {\ensuremath{{\mu{m}}}}
\newcommand{\Msun}    {\textrm{M}\ensuremath{_{\odot}}}
\newcommand{\Mt}      {\ensuremath{\textrm{M}_\textrm{tot}}}
\newcommand{\Rbrk}    {\ensuremath{\textrm{R}_\textrm{brk}}}
\newcommand{\Redge}   {\ensuremath{\textrm{R}_\textrm{e}}}

\renewcommand{\ge}    {\ensuremath{\geq}}
\renewcommand{\le}    {\ensuremath{\leq}}

\renewcommand{\sec}   {\ensuremath{^{\prime\prime}}}
\newcommand{\arcsinh} {\ensuremath{\textrm{arcsinh}}}
     
\begin{document}
\maketitle
\abstract{The circular velocities of the inner region of disk galaxies are predicted by standard physics but velocities beyond the stellar disks are not consistent with Newtonian physics if the material there is in stable circular orbits. However, this material is not gravitationally bound and so does not trace the gravitational field in the way that is usually assumed.
The gravitational attraction near the edge of a flattened mass distribution is significantly greater than that of an equal mass in a spherical distribution. The size of the effect depends on the specifics of the mass distribution but is greater than a factor of two for reasonable models. In fact, the circular velocity can exceed the escape velocity so that these galaxies are gravitationally unstable in way not previously considered and disk material is lost due to thermal escape, bars or other disturbances.
The nearly constant velocity observed in the outer disk region has been interpreted to mean that the dynamical mass of galaxies is much larger than the observed mass. In fact, there is no great discrepancy and no need to invoke dark matter at these scales. The gravitational field of a disk galaxy is determined at all radii by the observed mass. In the region of the stellar disk, stars and gas move in nearly circular orbits at velocities consistent with the gravitational field. In the outer regions the gravitational force drops rapidly so that stars and gas move outward almost unaffected by the attraction of the host galaxy.}

\section{Introduction}\label{sect:IsIntro}

The flat outer rotation curves of spiral galaxies have been interpreted to mean that the mass distribution of spirals is dominated by dark matter resulting in the currently accepted \LCDM\ model \cite{Pee17,Ber18,Sof01,Sal19}.
The concept was readily accepted due to the success of dark matter in explaining the distribution of the CMB.

However, the search for dark matter candidates has lasted decades and failed to find a particle which fits requirements and although the \LCDM\ model is successful at large scales it has severe problems at the scale of galaxies. \cite{Bul17} and \cite{Wei15} are recent reviews of these problems which \cite{San09,Kro12,Hag18} and others claim are severe enough to constitute a falsification of \LCDM\ at small scales.

In the 1950's Burbidge and others used the analytical solution for the oblate spheroid to model the rotation of the stellar disk and found that the resulting M/L ratios were reasonable (see especially \cite{Bur59}).  Subsequently \cite{Cas83,Kal83::2,Ken86,Buc92} showed that the velocity curves of the stellar disk are explained by a truncated disk model. Only after flat rotation curves were observed in \HI\ far beyond the stellar disk \citep{vanAlb85,vanAlb86} was it conclusively decided that dark matter was required and was in fact the dominant component of spiral galaxies.

The evidence for dark matter in early-type galaxies is scant. \cite{Rom03} found no dark matter in early-type galaxies although other studies give conflicting results. \cite{Bin08} explains the lack of dark matter by postulating that ellipticals, which are usually found in groups or clusters of galaxies, may have been stripped of their halos in interactions or their halos might have merged to form a single halo belonging to the entire ensemble. Also, \cite{Mil03} points out that for the most part the gravitational attractions probed by \cite{Rom03} are larger than Milgrom's constant and so the data is equally consistent with MOND and with standard Newtonian predictions assuming no dark matter.

The calculation of the required amount of dark matter  depends on the assumed mass distribution. The commonly used maximum disk model assumes that the baryonic matter is distributed in an exponential disk, extending to infinity. The amount of dark matter is that required to make up the difference between the assumed distribution and the actual distribution. 

Near-infrared observations provide more accurate M/L ratios compared to optical M/L ratios, eliminating the need for bounding assumptions. 
Calculations using NIR mass distributions and best-estimate M/L ratios reproduce stellar disk velocity curves with little need for dark matter \textbf{if} the measured distribution is used in a Poisson solution.
Recent studies (e.g., \cite{Yoo21,Cou15,Swa12})  confirm that the rotation curves of the stellar disk can be explained by applying reasonable M/L ratios to the observed flux but not the flat outer rotation curves of extended disks.
\cite{Bot02,Ang15,Li18} and others come to the same conclusions in the context of Modified Newtonian Dynamics (MOND): the velocity of the inner disk is consistent with the observed baryonic mass whereas the flat velocity of the outer region is explained by supposing that Newtonian gravity must be modified for low acceleration.

The mass distribution of the disk can also be estimated by employing the vertical dispersion of tracers and an estimated scale height. \cite{Ani18,Ani21} uses this technique  and corrects an error affecting several previous analyses which used inconsistent values of dispersion and scale height. 
The errors were large enough that dark matter had been thought to be required in the inner regions.
%(e.g., \cite{Ber11})
The results of this sort of analysis are now broadly consistent with mass distributions based on surface brightness measurements.

There are a number of alternate theories to \LCDM\ which try to explain the flat outer rotation curves of disk galaxies.
The most mentioned alternative theory  to \LCDM\ is the MOND hypothesis \citep{Mil83,Mil83::1,San02,McGau12} that proposes an ``effective" force law to explain the observed flat rotation curves of disk galaxies. Accelerations greater than Milgrom's constant, \aMil,\ are governed by Newtonian physics so that \aeff, the observed acceleration, is equal to \aN,\ the acceleration calculated using classical theory but accelerations $a < \aMil$ are governed by an effective force law
$ {\aeff = \sqrt{\aN \aMil}} $.  \aMil\ is thought to be a universal constant with a value of \aMil $\approx$1.2 \AA~s$^{-2}$.
The Dark Matter disk (DMD) \citep{Syl23} model assumes that dark matter is confined to the stellar disk and the distribution of the dark matter follows the distribution of the observed gas.
Other explanations include the magnetic hypothesis \citep{Bat02}; Verlinde's ``Emergent gravity" theory \citep{Ver17}; and a several attempts to revise general relativity (e.g., \cite{Ras72}). 
None of the proposed alternate solutions has been generally accepted and the need for a new paradigm is apparent \citep{Nes23}. 
\subsection{Organization}
This paper is organized as follows: \\
 Section \ref{sect:AreTruncated}: 
    Properties of stellar disks are discontinuous at the radius where the gravitational attraction of the disk becomes negligible and the disks are truncated beyond this radius. \\
 Section \ref{sect:TruncDisks}:
    Exact analytical results for the potential and gravitational fields for a few idealized disk models show that neglecting the effect of disk truncation results in an error of 2 to 4.\\
 Section \ref{sect:NGC3198}:
    Re-evaluation of NGC 3198 shows that basic misunderstandings affected the analyses of this much studied galaxy.\\
 Section \ref{sect:Discussion}:
    Stellar disks of late-type galaxies are unstable in the region just beyond the edge in a way not previously considered.
    Realization of this fact explains the many problems with the standard CDM model.\\
% Section \ref{sect:WhatWhy}: A brief discussion of the impediments which slowed progress for so long.\\
 Section \ref{sect:Summary}:
    Summary and conclusions.
\section{Galaxy Stellar Disks are Truncated }\label{sect:AreTruncated}

Here I discuss recent work which shows that stellar disks are not, in general, exponential to great distances but rather are truncated just beyond the Holmberg radius.

 The maximum disk model \citep{Fre70,Bin08} holds that stellar disks extend great distances with a single exponential scale. However, studies of inclined galaxies found that the stellar mass distribution often follows a smooth exponential to a breakpoint where:\\
1) Some galaxies continued exponentially with the same scale factor;\\
2) Some (called truncated) continue but with a smaller scale factor; and\\
3) Some (called anti-truncated) continue but with a larger scale factor \citep{Erw05,Poh06,Poh08}.

In contrast, \cite{vanderKru07,vanderKru08} found that the majority of edge-on spiral galaxies show evidence for complete truncation of the stellar disk. Truncations occur somewhat beyond the Holmberg radius and in many cases, the truncation radius is associated with an abrupt change in rotational velocity and with large \HI\ warps. \cite{vanderKru07} makes the point that the transition from the flat inner disk to the warped outer disk is {\it abrupt} and {\it discontinuous} and the inner flat disk and outer warped disk are distinct components.

Some late-type galaxies have a significant amount of gas, stars, and even ongoing star formation in the outer disk, often associated with spiral arms or recent disturbance and both gas and stars can be seen at great distances beyond the inner disk. For instance, \cite{Lop18} found young disk stars in the flared outer disk of the Milky Way at R $>$ 26 kpc and even at R  $>$ 31 kpc, far beyond the edge of the inner disk usually taken to be \app12-15 kpc (see also \cite{Car10}). The metallicity distribution of these stars in the Galactic plane is distinct from that of the halo populations but consistent with the stars of the inner disk, indicating some sort of inside-out formation or outflow. Similarly, \cite{Jan20} recently studied the radial mass distribution of NGC 300 with archival HST data.  Several populations of bright stars were imaged reaching much further and deeper than previous studies. The observations found a downward (type 2) bend just beyond the visible edge of the stellar disk (at $\mu_{3.6} >27 \mas$) and found an upward (type III) bend at about 9 kpc beyond which the mass distribution continues as exponential out to at least 19 kpc. The age of the stellar population transitions from primarily young stars in the inner disk to entirely old ($\ge$ 9 Gyr) RGB stars in the outermost region.

\cite{Bak12}, \cite{Mar14} and \cite{Kna17} reconcile the observations of edge-on and inclined galaxies by showing that the stellar haloes of disk galaxies can outshine the brightness of inclined galaxies near the edge of the disk so that beyond the truncation radius the observed stellar density falls off exponentially but the scale factor is that of the halo and does not, in general, describe the disk proper. Similarly, \cite{Pet17} find that the presence of truncations and of extended stellar halos are mutually exclusive which implies that the presence of a stellar halo and/or light scattered by the point spread function can hide truncations and so very extended exponential disks are not as common as previously thought. \cite{Mar19} studied the truncations in two nearby (D\app15 Mpc) Milky Way-like galaxies in NUV, optical and 3.6 \mm\ wavelengths and found that, at all wavelengths, the truncation occurs altitudes as high as 3 kpc above the mid-plane. The stellar mass density at the position of the truncation is \app1-2 \Msun\ pc$^{-2}$, consistent with typical star formation thresholds, and the U-shaped radial colour profile seen at the location of the truncation indicates a star formation threshold.

To be sure, complete truncations are not universal and in some cases the outer disks extend to great distances.
Here, the word ``truncation" is used to mean the abrupt boundary of the gravitationally significant portion of a stellar disk. The surface density beyond this truncation radius is low (i.e., below \app 1 \Msun\ $ pc^{-2}$) and because of the flaring and warping of the outer disk the material  must be treated as part of the halo in calculating the force field, making a negligible contribution. In this sense, all stellar disks are truncated and this is important because the gravitational field at the edge of a truncated disk is much larger than that of a sphere or an infinite exponential disk with the same mass.

 \cite{GildePaz17,Kna17} provide new detail of the morphology of the outer region of disk galaxies. This region is distinct from the stellar disk and properties change abruptly at the interface.
The structure of the disk changes at the edge, flaring and bending of the disk occur and colour gradients are different in the outer region. Truncations or `breaks' are seen at the interface and beyond this radius the rotation curve is not consistent with the observed disk mass. 
Most recently, \cite{Cha22} used deep imaging (to $\mu_g =29.1 \mas; \mu_r=28.5 \mas$) of a set of \app1000 low-inclination galaxies to find a distinct cutoff of the stellar disk at the radius where star formation drops precipitously. Beyond this radius the star formation rate is seen to fall off dramatically. 

The  quantities of galaxy dynamic quantities are related by four correlations which together confirm the idea that the stellar disks are finite.
The most well-known of these correlations, the Baryonic Tully-Fisher (BTFR) relation \cite{Tul77,Fre99,McGau00::1,McGau05::1}), correlates total baryonic mass, including both stellar mass and gas, to rotational velocity measured resulting in a tight relationship over a range of many decades of mass. 
\cite{San20,Sch17,She23,Tru20}) studied three correlations of galaxy mass, size and velocity and found that the scatter of the correlations is much reduced when: 
\begin{itemize}
  \item \Rbrk\ is used as the measure of size where  \Rbrk\ is the radius of the edge of the stellar disk
  \item \Mt\ is used as the measure of galaxy mass where \Mt\ includes only the baryonic mass
\end{itemize}
which supports the idea that disks are truncated and either consist only of baryonic mass or else the dark matter is strongly coupled to the baryonic mass.

\section{Properties of Truncated Disks} \label{sect:TruncDisks}
Late-type galaxies are typically quite flat with the ratio $h_z/h_R < \app0.1$ and in some cases as small as 0.01. These galaxies can be modelled as flat disks when calculating the gravitational field since the error compared to explicit 3-D calculations is negligibly small. 
\cite{Mes63} argues that it is possible to model spiral galaxies as one of two analytic flat finite disks. The first model, the Maclaurin disk, has a linearly rising rotation curve and the second, the finite Mestel disk (FMD), has a flat rotation curve over the entire disk, indicating a singularity such as a black hole at the center.
Observations of late-type galaxies are, for the most part, consistent with Mestel's proposition except that a late-type galaxy with a prominent bulge must be modelled with several components.

Although it's well known that the gravitational attraction of a flat disk is greater than that of a spherical model \citep{Bin08}, the size of the effect is not widely appreciated.  As the simplest example, the force due to a thin disk with constant surface density increases with radius and then spikes to infinity at the edge of the disk \citep{Mes63,Las83}. More realistic models show a less dramatic increase at the disk edge but, even so, the effect is on the order of a few which is enough to explain the rotation curves of stellar disks.

In contrast, the attraction of an exponential disk is greater than that of a spherical model in the outer regions \citep{Bin08} but the effect is small and covered by an allowance of about 10-20\%.     % figure 2.17
\subsection{The Maclaurin Disk}

\begin{figure}
\includegraphics[width=0.95\linewidth]{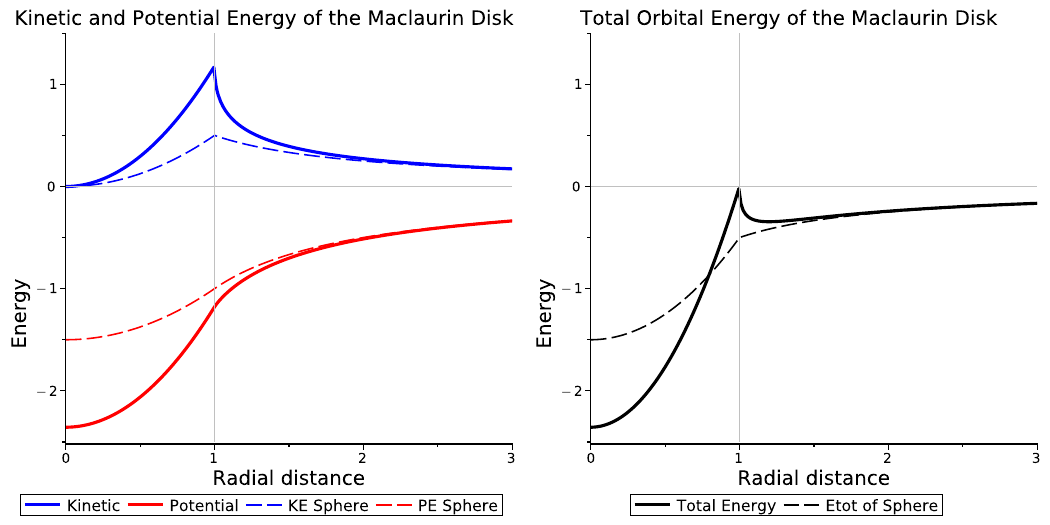}
 \caption{(a) Kinetic and Potential Energy of the Maclaurin disk compared to a homogenous sphere with the same mass and radius. For the sphere, the ratio KE/PE = -0.5 for external points. For the disk, the ratio KE/PE = -1.0 at the edge and quickly approaches -0.5 for larger radii. (b) The total energy of a particle orbiting in the field of a Maclaurin disk spikes near the edge and quickly drops just beyond the edge.}
\label{fig:OrbitalEnergyMaclaurin}
\end{figure}

Beginning in the early 18th century Colin Maclaurin, along with James Ivory and many others, studied the properties of elliptical bodies. \cite{Cha87} includes a very good historical summary. See also \cite{MacMil30,Bin08,Ber00,Schm56,Mih68,Kal71,Kal72}. The homogeneous oblate spheroid is the simplest case of a spinning body for which the gravitational attraction balances the centrifugal force.

The Maclaurin disk, also known as the Kalnajs disk \citep{Mes63,Kal72,Sch09}, results if an oblate spheroid collapses along the minor axis. This disk  has the mass distribution:
    \begin{equation} \label{eq:SurfDensMac}
          \Sigma_{Mac}  =   \frac{3 M }{2 \pi \Redge^3}\sqrt{\Redge^2-R^2}  ,~~~~ R \le \Redge
    \end{equation}
and is a good approximation of late-type bulgeless spirals.

The orbital velocity rises linearly to the edge of the disk and then decreases rapidly to become identical to that of a spherical mass. At the edge of the disk, the gravitational attraction is a factor of  $ {3\pi}/{4} \simeq 2.36$ larger than that of an equivalent sphere and so the rotational velocity is $\sqrt{2.36} = 1.54$ times greater. Figure \ref{fig:OrbitalEnergyMaclaurin} compares the kinetic and potential energy of a Maclaurin disk to that of a homogenous sphere with the same mass and radius.  The kinetic energy of a particle orbiting near the edge is larger by the factor of 2.36 compared to that of a particle in orbit around a sphere of equal mass. The difference in potential energy is less dramatic because potential energy is largely determined by the force field at greater distances. The result is that the kinetic energy of a particle in orbit at the edge of a Maclaurin disk is equal to the potential energy; i.e. the orbital velocity is equal to the escape velocity and if a particle moves outward from the edge disk it continues out to infinity.

Notice that the total orbital energy \emph{decreases} with radius in the region which extends from the edge, to  \app1.20 \Redge.
A particle in this region with velocity $V = \sqrt{R*F}$ is in an \emph{unstable} orbit.
If the velocity of the particle is disturbed slightly would move outward to a stable orbit.
Stars and gas in this region of instability are just passing through: moving outward from the disk to eventually join the galaxy halo or leave the galaxy entirely.
This edge effect promotes outflows and impedes the predicted (but unobserved) inflows from the cosmic web.

\begin{figure}
\includegraphics[width=0.95\linewidth]{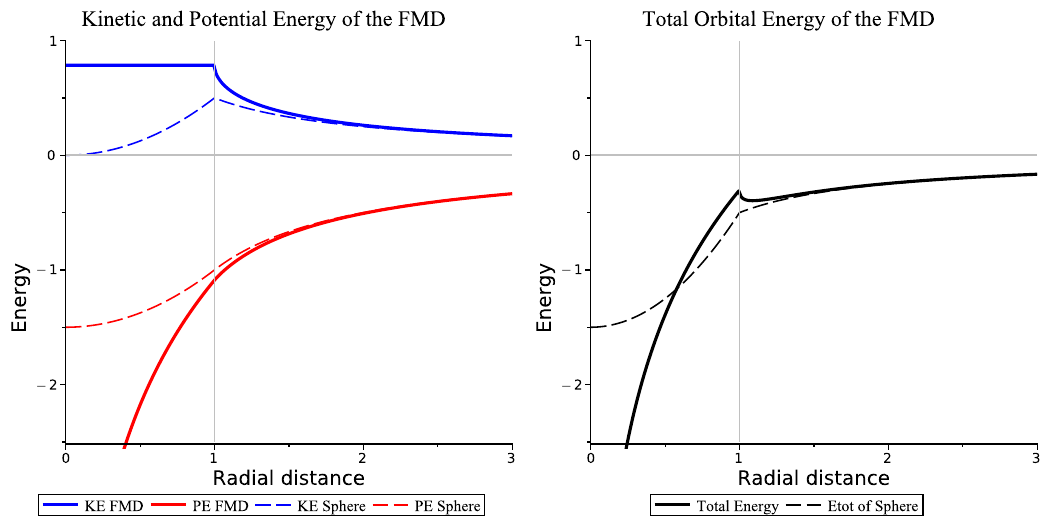}
\caption{(a) Kinetic and potential Energy of the Finite Mestel disk compared to a homogenous sphere with the same mass and radius. The kinetic energy spikes at the edge of the disk but not as dramatically as seen in Figure \ref{fig:OrbitalEnergyMaclaurin}. (b) The total  energy of a particle orbiting in the field of an FMD drops just beyond the edge of the disk, creating a region of instability.}
\label{fig:OrbitalEnergiesFMD}
\end{figure}

\subsection{Finite Mestel Disk}

The Finite Mestel disk (FMD) \citep{Mes63,Sch12}, has the mass distribution:
\begin{equation} \label{eq:SurfDensFMD}
 \Sigma_{\FMD} =   \frac{M }{2 \pi \Redge R}  \arccos\left(\frac{R}\Redge\right) ,~~~ R \le \Redge
\end{equation}

The mass distribution of the FMD includes a central cusp where the rotation curve steps to a value which is then constant to the edge of the disk. Beyond the edge of the disk, the orbital velocity drops much more quickly than Keplarian and approaches the $1/\sqrt{R}$ curve asymptotically.
The flat velocity is typical of many Sc and Sd galaxies where a central bulge, and often a massive black hole, approximate the singularity.

Figure \ref{fig:OrbitalEnergiesFMD} shows the energy measures of a particle orbiting in the field of a FMD.  At the edge of the disk, the attraction is a factor of  $ {\pi}/{2} \simeq 1.57$ larger than that of a sphere with the same mass and so the rotational velocity is $\sqrt{1.57} = 1.25$ times greater.

Like the Maclaurin disk, the total orbital energy of the FMD decreases with radius in the region just beyond the edge of the disk where stable orbits are impossible. The effect not as strong however, so that spontaneous escape from the disk is expected to be less energetic and might be limited to a flow of the warmest gas.

\subsection{Potential of a homogeneous bar}

\begin{figure}[t]
\centering
\includegraphics[width=0.7\linewidth]{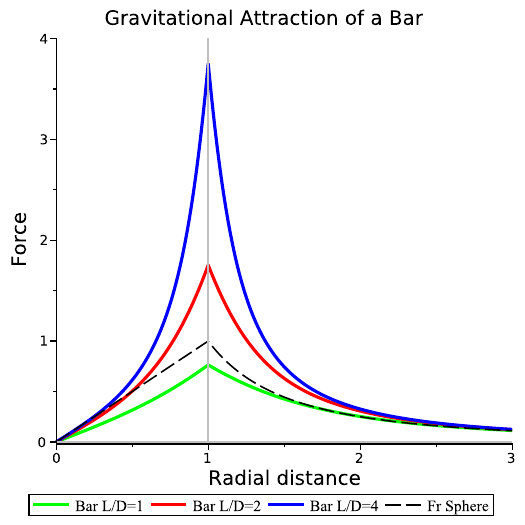}
\caption{The force at the end of bars of equal mass and length but different diameters increases with decreasing diameter as approximately $L/D$. The force of a homogeneous sphere of equal mass is shown for comparison.}
\label{fig:Bar1-Fr}
\end{figure}

\begin{figure}[ht]
\includegraphics[width=0.95\linewidth]{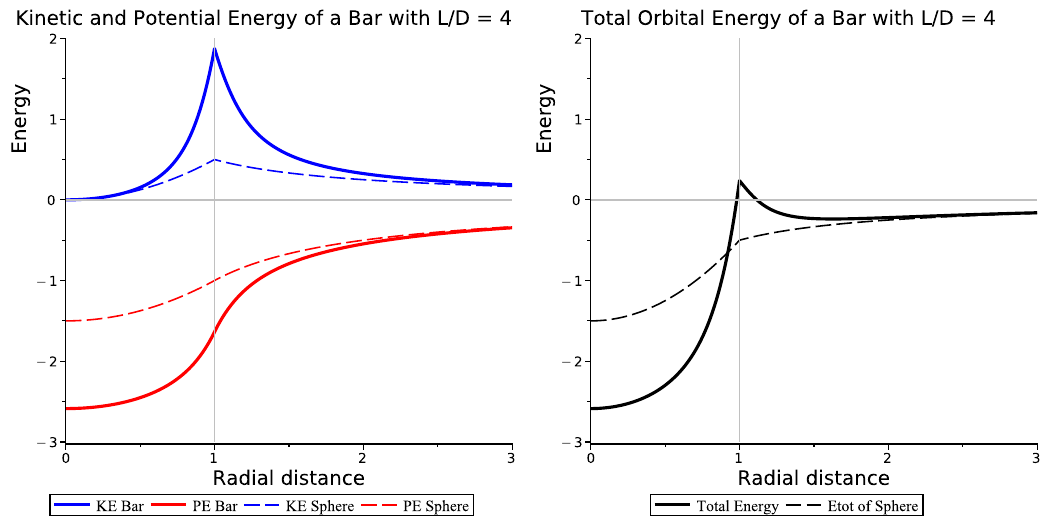}
\caption{(a) Kinetic and Potential Energy of a bar compared to a homogenous sphere with the same mass. The sphere has radius 1 and the bar has a total length 2 and diameter 1/2. The force and potential of the bar are calculated along the axis. (b) The total  energy of a particle orbiting in the field of the bar spikes near the end and quickly drops just beyond the edge.}
\label{fig:OrbitalEnergyBar}
\end{figure}

Galaxy bars are common in the nearby universe: about 70\% of spirals show bars in the near infrared \cite{Esk00}.
These bars vary greatly in complexity and appearance.
Some completely span the stellar disk, dominate the dynamics of the disk and are often associated with grand-design spiral arms.
Others extend only part way toward the outer disk and may appear to be flat.
Still other bars consist of two halves joined together at the galaxy center by a peanut-shaped bulge, likely the result of buckling \citep{Ath16}.
Bars seem to evolve secularly, becoming weaker, shorter and relatively faster rotating. Some galaxies have two or even three bars of different lengths suggesting a process of sequential bar formation and destruction.

The Appendix provides the solution for the gravitational force and potential along the axis of a homogeneous cylinder. This simple model is sufficient to illustrate some important properties of bars.
Figure \ref{fig:Bar1-Fr} shows the force of a bar for several values of the quantity L/D: the ratio of the total length to diameter which is one of several measures of bar strength (cf. \citep{Agu99,Gar17}). The attraction at the end of the bar is seen to vary inversely with the ratio for bars of the same mass. 

A value of approximately 4-5 for the L/D ratio is the maximum seen for strong bars and I adopt a value of L/D=4 so that the attraction is a factor of $\simeq 3.75$ larger than that of a sphere with the same mass and the rotational velocity is $\sqrt{3.75} \simeq 1.9$ times greater.

Figure \ref{fig:OrbitalEnergyBar} shows the energy of a particle moving in the field at the end of this bar. The kinetic energy of a particle streaming across the bar is larger by the factor of 3.75 compared to a particle in a circular orbit around a sphere with the same mass. The difference in potential energy is less dramatic so that \Etot\ is significantly positive at the end of the bar. That is, the velocity which balances the gravitational attraction is greater than the escape velocity. The radial derivative of \Etot\ is negative in the region just beyond the end of the bar so that stable orbits aren't possible in this region.

\subsection{Comparison of several mass distributions}
\setlength{\tabcolsep}{4pt}
\begin{table}[ht]
\centering
\caption{CAF Factors for Several Models}
\begin{tabular}{ l l l}
\hline
~~~Name &Surface Mass Density                                                                          & \CAF      \\[+2.0ex]
   Finite Mestel Disk & $ \Sigma = \frac{\Mt}{2 \pi \Redge^2 } \, \arccos\left(\frac{r}\Redge\right)  $&$  1.57  $ \\[+2.0ex]
   Maclaurin disk     & $ \Sigma = \frac{3 \Mt}{2 \pi \Redge^2} \sqrt{1 - {r^2}/{\Redge^2}}           $&$  2.36  $ \\[+2.0ex]
   Solid Bar          & $ \textrm{Assumed  }  L/D = 4                                                 $&$  3.75  $ \\[+2.0ex]
   Exponential Disk   & $ \Sigma = \frac{\Mt}{2 \pi h^2} \exp{-\frac{r}{h}}                           $&$  0.83  $ \\[+2.0ex]
\hline
\end{tabular}
\label{tbl:FlatFact}
\normalsize\begin{minipage}{0.95\linewidth} Augmentation factors (\CAF) for four models of disk galaxies vary by a factor as large as 4-5. The \CAF\ factor is the increase in gravitational force compared to a point mass calculated at the edge, except that \CAF\ of the exponential disk is calculated at the radius of maximum velocity.  \end{minipage}
\end{table}

Here I define a parameter \CAF\ to compare the strength of several mass distributions. \CAF\ is the augmentation factor which describes the increase in gravitational force compared to a spherical distribution and is defined by
$$\CAF = A_{disk}(R_d)/A_{point}(R_d)$$
where $A_{disk}(R_d)$ is the acceleration at the edge of  some mass distribution and $A_{point}(R_d)$ is the acceleration of a point mass with the same total mass: $A_{point}(R_d) = M_t G / R_d^2$.
Table \ref{tbl:FlatFact} gives \CAF\ for a few simple models in a format to allow cross-comparison.
\CAF\ ranges in value from less than 1 for an exponential distribution to about 4 for a strong bar.

The augmentation factors for truncated mass distributions are large and ignoring these effects has lead to large self-perpetuating errors.

\section{Re-evaluation NGC 3198}   \label{sect:NGC3198}

\begin{figure*}[ht]
\centering
\includegraphics[width=0.8\linewidth]{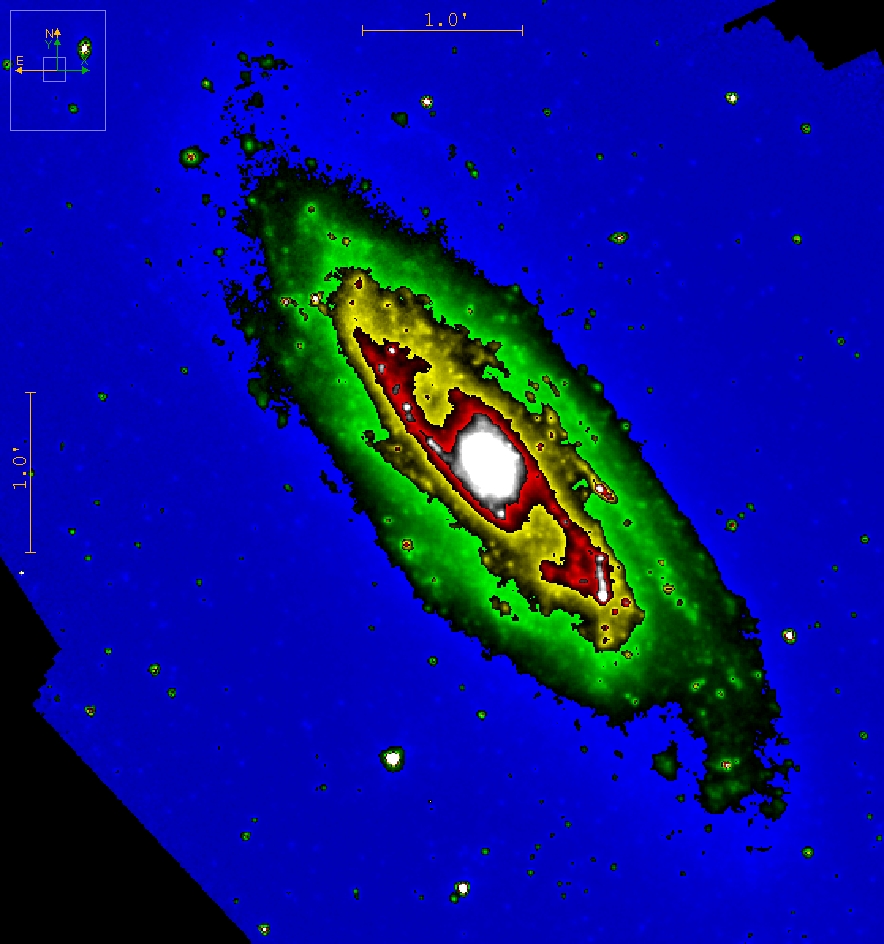}
\caption{False colour Spitzer 3.6 \mm\ image of NGC 3198 shows the inner disk which rotates in the counter-clockwise direction. A boxy bar dominates the kinetics and significant .}
\label{fig:3198Spitzer-5ColorBar}
\end{figure*}

NGC 3198 is the prototypical disk galaxy which convincingly demonstrated the need for dark matter to explain the rotation curves of spiral galaxies \citep{vanAlb85}. However, almost 40 years have passed and there has been very little progress in understanding the phenomenon.  Observational and analytical tools have improved greatly during this time and here I revisit the dynamics of the NGC 3198 using current data and models.

Figure \ref{fig:3198Spitzer-5ColorBar}  is an image produced from an archived 3.6 \mm\ Super Mosaic image from the Spitzer Enhanced Image Program(SEIP).

This NIR band image traces stellar mass which is by far the most important baryonic contributor to the gravitational field of NGC 3198.
To generate this figure, the bright central region was saturated and equal spaced colour ramps were applied to disclose the details of the mass distribution.

\begin{figure}[t]
\centering
\includegraphics[width=0.76\linewidth]{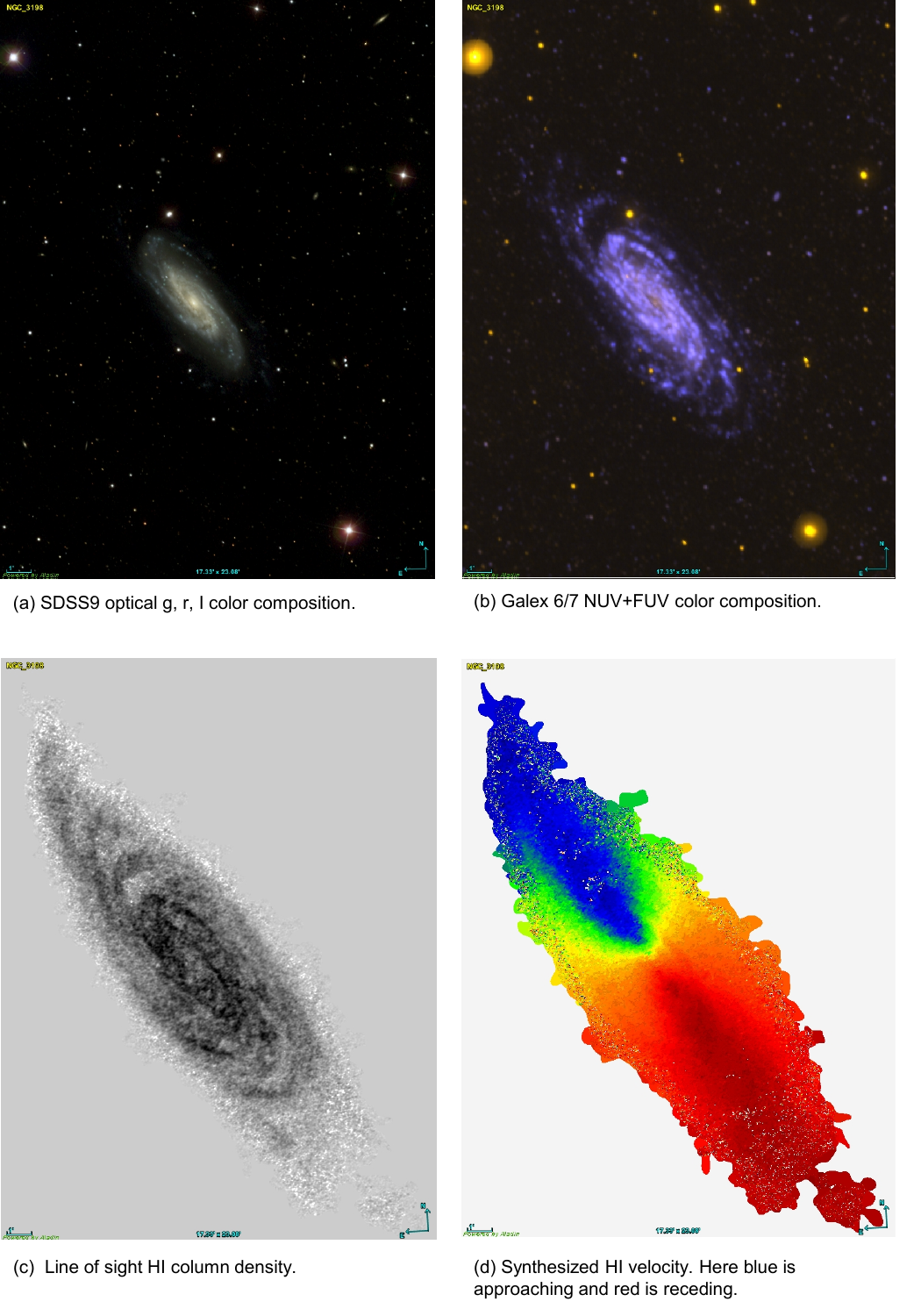}
\caption{Four views of NGC 3198 in various wavelengths from archived data, all to the same scale.    These views show that gas and stars have separated from the inner disc and now form a rapidly expanding outer disk. The attraction of the light inner disk doesn't slow the ejecta appreciably so that we observe a nearly constant azimuthal velocity in the outer disk. }
\label{fig:3198fourfigs}
\end{figure}

Figure \ref{fig:3198fourfigs} shows four views of NGC 3198:
Figure \ref{fig:3198fourfigs}.a shows NGC 33198 in SDSS colourized optical which was  accessed in Aladin.
Figure \ref{fig:3198fourfigs}.b is a Galex colourized ultraviolet image which  was accessed in Aladin.
Figure \ref{fig:3198fourfigs}.c and Figure \ref{fig:3198fourfigs}.d show 21-cm \HI\ intensity and velocity data generated for the SINGS project \citep{Ken03}.
The four figures were aligned and cropped in Aladin.

The figures use a common scale length which is smaller than that of Figure \ref{fig:3198Spitzer-5ColorBar} by a factor of \app3 so that Figure \ref{fig:3198Spitzer-5ColorBar} covers only the innermost region of these images.
The images in Figure \ref{fig:3198fourfigs} are not new but are important to show scale and especially to point out the branching trajectories seen in \ref{fig:3198fourfigs}.c which are incompatible with a standing density wave model.

Previous studies \citep{Hun86,Gen13,Tra08} dismissed the possibility that NGC 3198 has a significant bar but the good detail of the Spitzer image reveals an evolved bar extending almost to the outer edge of the stellar disk and this structure contains most of the system mass and dominates the dynamics of the galaxy.

The bar rotates counter-clockwise and the surface density of the leading edges of the bar is greater than that of the trailing edges by a factor of about 2.
The thickened leading edges are offset from the centerline of the bar to form a shape which is uncommon but not unique. For instance, NGC 4293 shows a similar structure in NIR \citep{Que21}.

Velocity-position data in both optical and \HI\ \citep{Hun86} show that velocity increases linearly with radius almost to the end of the bar, indicating solid-body rotation, so that the bar/bulge structure is most likely long-lived. The constant velocity region begins at the end of the bar where material separates outwardly.

\clearpage

\clearpage

The central bulge includes a very bright small sphere that flattens to the boxy bulge joining the two segments of the bar. This is the black hole with a mass of \app2.5\dex{6} \Msun\ reported by \cite{Don06}.

Material is seen to be thrown from both the Southwest and the Northeast sides and these outflows are the origin of the optically prominent spiral arms. A number of spherical clusters can be seen in the outflow. The largest (on the Southwest side) has a magnitude of 17.6 AB mag for a mass of 2.2\dex{7} \Msun. As noted in \cite{Hun86}, these `alpha knots' can be seen out to the Holmberg radius.

Spurs along the bar indicate that there is \emph{outward} flow along the leading edges of the bar, some of which continues outwardly and joins the spiral arms. There is \emph{inward} flow along the trailing edges toward the central bulge. The result is a circular flow pattern around the bar.

I used the bar model of the Appendix to calculate the force and potential along the axis of the bar of NGC 3198. The chosen model is the simplest analytic approximation for the radial force of a bar and the solution involves only elementary functions.  The adopted data for the calculation are:

\begin{itemize}
\item The distance of NGC 3198 is 13.8 Mpc \citep{Kar15} so that \mbox{$ 1 \sec  \simeq   66.9 $ pc} and the length of the bar is 8.2 kpc.
\item The the 3.6 \mm\ magnitude is 10.315\ AB mag \citep{Sor14} so that the total luminosity is L=3.66\dex{10} \Lsun. Assuming a mass to light ratio of $\Upsilon_{3.6} =  0.6$ \citep{Mei14,Sch17} the total stellar mass of the system is approximately 2.2\dex{10} \Msun.
\item Assuming an inclination of 0.7 \citep{Gen13}, the width of the bar is \app3.0 kpc giving a L/D ratio of 2.7.
\item The gas mass is  \app20\% of the total galaxy mass but is mostly beyond the inner disk, is flared, and doesn't contribute significantly to the force field and this components is neglected.
\end{itemize}

\begin{figure}[t]
\centering
\includegraphics[width=0.7\linewidth]{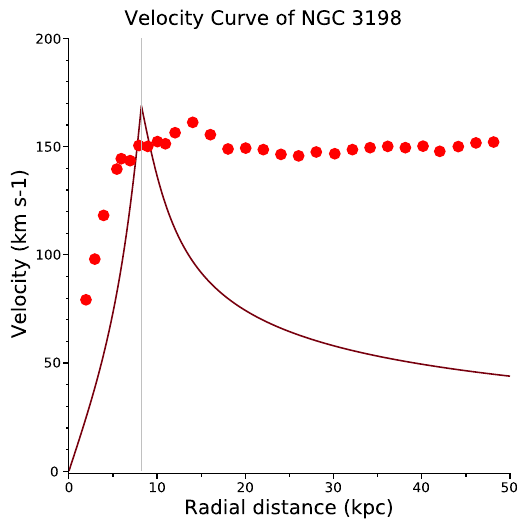}
\caption{The calculated rotation curve based on a stellar bar with a length of \app8.2 kpc (2.0 arcmin) is compared to the recent observed data. \Redge, the edge of the stellar mass is indicated by the vertical grey line.}
\label{fig:NGC3198Result}
\end{figure}

The velocity curve was calculated as $Vc = \sqrt{ R Fr}$ where the $Fr$ is the radial force at a distance $R$ from the center of the galaxy at the centerline of the bar.
Figure \ref{fig:NGC3198Result} shows the resulting velocity curve compared to the measured data given in \cite{Kar15}.  The calculated velocity explains the measured data reasonably well for $R<\Redge$ without the need for dark matter.

The difference between the calculated and measured velocities in the region beyond the end of the bar is profound and can't be explained by any plausible dark matter halo, which would need to be hollow for $R<\Redge$, rapidly spike to very high density at \Redge\ and then quickly decrease to the density profile that results in the observed flat rotation curve. An attempt to reconcile the observations in MOND would involve similar absurdities and it is seen that both these theories depend on the non-physical assumption of an infinite exponential mass distribution for the stellar disk.

The only reasonable alternative is that the gravitational force of the inner galaxy is determined by the observed mass and the observed outer rotation curve does not measure circular motion but rather the azimuthal velocity of gas leaving the system almost unimpeded by the low mass of the galaxy.

This re-evaluation demonstrates how the methods used to determine rotation curves have gone badly astray. Fitting a symmetric equilibrium tilted ring model to observations of a non-symmetric non-equilibrium case produces a plausible rotation curve which is completely wrong and this sort of error, repeated many times, has resulted in the current impasse.

\section{{A new view of galaxy dynamics}}  \label{sect:Discussion}
\emph{ ``The high velocity of rotation argues that in some cases, at least, -as NGC 4594, for instance- the nebula is, in consequence of rotation, expanding. Indeed, the disk form and the spiral arms of these nebula imply action past or present, of expansive forces. The evidence from these observations, and from other sources, to my mind, makes clear the need of our entertaining the view that systems exist which are undergoing expansion."(Vesto Slipher, 1917)}

A dynamical system will evolve to reach the lowest energy configuration if there is a mechanism to do so. In the case of a rotating disk, the lowest energy state occurs when all the mass is concentrated at the center and all the of the rotational inertia is carried by a massless particle at infinity and we do, in fact, observe that disk galaxies throw matter outward while the remaining mass concentrates in the central region.

There are two sorts of radial outflows which drive galaxy evolution:
\begin{itemize}
\item %{As shown in Section \ref{sect:NGC3198}}
   Grand design spirals in the outer disk regions are outflows caused by galaxy bars or other drivers and NGC 3198 is a fine example of the process. These spirals very dramatically drive material outwards, cooling and compressing the host galaxy. A central black hole is often seen which accretes inflowing matter.
\item Gas and stars can be lost continuously from the edge of an extended flat disk by the thermal escape mechanism. Kinetically hot material is lost preferentially, cooling the galaxy and eventually leading to a razor-thin, almost gasless low surface brightness galaxy surrounded by an expanding gaseous disk.
\end{itemize}

These radial flows are qualitatively different from the vertical gas outflows from innermost regions of large galaxies. The vertical outflows consist of high velocity winds and fountains powered by star bursts and active galactic nuclei \citep{Vei05,Vei20,Zha18}.
Outflows from the edge of a disk galaxy are not so easily recognized.  A particle separating from the disk continues with almost the same azimuthal velocity and direction as a bound particle. The initially small radial component of velocity increases as the body travels outward following a parabolic path..

\subsection{Spiral Outflows}
\cite{Too77,Dob14,Sel22}  review of the theory of spiral structure.
The ubiquity of spirals has been taken to mean that they are long-lived (on the order of \app 10 revolutions) but this leads to the winding problem.
The density wave theory \citep{Lin64,Lin66,Ber96}  holds that spirals are rigidly rotating static patterns caused by a gravitational instability of the stellar disk. This theory avoids the winding problem but is beset by other fundamental issues. In particular, the instability criterion for spiral formation is not satisfied in the outer disk regions where grand design spirals are seen.
An alternate theory of a recurrent cycle of groove modes \citep{Sel14,Sel19} has emerged but it has been difficult to apply this theory beyond the bright inner disk \citep{Kho15}.

Recent studies suggest find that grand-design spiral arms are episodic material flows rather than density waves. \cite{Gue11,Bab15,Wad11} investigated spiral patterns in standard \LCDM\ cosmological models assuming massive dark halos. These studies cast doubt on the standard theories in two ways. First, only transient material arms were seen, rather than long-lived patterns. Second, massive cuspy dark matter halos result in relatively light disks which in turn produce only flocculent multi-arm structures \citep{Don16}. To date there have been no \LCDM\ simulations which reproduce the spirals as we observe them.

There is ample observational evidence for the idea that grand design spirals are fundamentally different from the spirals seen in the stellar disk. For instance \cite{Ken15} analysed spiral structure in a sample of galaxies from the \emph{Spitzer} Infrared Nearby Galaxies Survey (SINGS) and found that either a tidal interaction or a strong bar is a necessary condition for driving grand-design spiral structure. Also, galaxies with strong bars tend to have more loosely wound arms than unbarred spirals and \cite{Mas19} takes this as evidence that the presence of a strong bar might prevent winding.

\cite{Ben17,Ben19,Syl20} report the results of numerical simulations of the collapse of gas clouds into flat galaxies under influence of strictly Newtonian gravity without dark matter. Extended spiral patterns occur which are similar to observed grand design spirals and consist of material moving radially outward. In addition, the simulations predict rings and bars which are similar to observations. This behaviour is suppressed in systems which include massive dark halos. 
\cite{Syl19} discusses the implications of the simulations: the dynamical mass of disk galaxies has been greatly overestimated, resulting in a perceived need for dark matter.

Figure \ref{fig:3198Spitzer-5ColorBar} shows mass flowing up the leading edge of the strong bar of NGC 3198. This flow acquires angular momentum from the bar and a portion of the flow travels outward from the inner disk and forms a spiral in the outer disk while retaining a high tangential velocity. Mass flowing down the trailing edge of the bar gives up angular momentum to the bar and a portion of this flow is absorbed into the bulge to fuel the central black hole. Potential energy is converted to radiation as the black hold acquires mass.
This bar/spiral mechanism is an efficient way for the system to reach a lower energy state.
The angular velocity slows with 1/R as the material moves outward, forming a trailing spiral.
The process continues until the bar has shortened and weakened enough that mass is no longer lost from the system.
After this, the spiral arms spread outward and fade to invisibility.

The two-armed spiral pattern of the outflow varies with wavelength and is most distinct in the visible spectrum. The stellar arms are typically seen to extend approximately 180 $\deg$ from the ends of the bar whereas the \HI\ streams are more extended radially and are less distinct (See Figures \ref{fig:3198fourfigs}:a and \ref{fig:3198fourfigs}:c ).
\subsection{Steady Symmetric Outflows}
The radial derivative of total orbital energy is \emph{negative} just beyond the edge of a late-type disk and this has profound implications.
A particle - a star or a hydrogen atom - which enters this region is no longer bound to the stellar disk and travels outward to an orbit in the outer disk region or might leave the system entirely.  Kinetically hot material is preferentially lost so that the disk is cooled and flattened and stellar orbits become more circular. The disk becomes more rotationally supported rather than pressure supported.
The evolution to a cooler, flatter disk continues unless it is interrupted by some disturbance such as an encounter with a nearby galaxy or spontaneous bar formation. The endpoint of the process is a cold, very flat, low surface brightness (LSB) disk surrounded by a vast expanding disk of escaping gas.
Such LSB galaxies are quite common but the way they form has not been well understood \citep{Bot97::1,Dal97,ONe00,Scho01}.

The model of steady radial outflow from an unstable disk addresses some of the anomalies related to LSB galaxies:
\begin{itemize}
 \item Galaxies  modelled in a \LCDM\ framework are much too hot kinetically \citep{Pee22}. That is, the stellar velocity dispersion in the plane of a disk galaxy is predicted to be much higher than what we observe. The problem comes about because there is no simple way to cool the stellar disk in the \LCDM\ simulations resulting in elongated orbits and  'puffy` disks.
 \item `Superthin' galaxies (STGs) are a subset of LSB galaxies which have major-to-minor axial ratios of about 10-20 and are typically found in isolated environments. \cite{Adi21,Nar22} find that superthin disks are LSB galaxies seen edge-on and are extremely cold kinetically, especially in the outer regions. \cite{Adi21} notes that the dynamical stability of these disks is a mystery. STGs don't arise naturally in \LCDM\ and speculative explanations (e.g., \cite{Haf22}) lack observational support but an STG is the obvious endpoint of a disk galaxy undergoing outflow of its hottest material.
\item Gas and stars originating in the disk can be seen in the outskirts of seemingly undisturbed LSB galaxies, indicating inside-out formation. This contradicts the usual CDM models of galaxy formation and is hard to account for in the presence of a massive dark halo which would stabilize galaxies against outward migration
\end{itemize}

\subsection{Gas Accretion}
Gas accretion onto the stellar disks of galaxies is necessary because without an external source of gas  most galaxies would run out of gas within a few Gyrs. The details of how this gas is acquired are still unclear. \cite{Bla17,Loc17,Put17,Pez16} summarize the situation to date.

\LCDM\ simulations predict that cold \HI\ filaments bring primordial gas to galaxies which enters the disk radially. However, very sensitive measurements using the Robert C. Byrd Green Bank Telescope (GBT) with detection limits as low as 6.3\dex{17} cm${-2}$, have failed to find the predicted filamentary flows \citep{Pis14,Pin18,Pis18,Sar21}. Although \HI\ is sometimes seen at great radially limits the gas is not is not dense enough to sustain the predicted flows. The search for filamentary flows continues with ever more sensitive measurements but it is unlikely that the predicted flows will be found. In any event, the potential barrier at the edge of disk galaxies described in Section \ref{sect:TruncDisks} prevents gross inward flows so that the \LCDM\ model cannot be correct in this respect.

The dominant accretion path is that incoming gas enters the extended hot \HI\ halo which envelopes most disk galaxies and then enters the disk.
Hot \HI\ gas cools slowly in isolation and so it had been thought that the gaseous halos of spirals could not be the source of the required \HI.
However \cite{Fra17} describes the process where cool gas and metals are thrown into the lower halo via galactic fountains and greatly enhance cooling.
Cooled gas then falls onto the disk in quantities great enough to sustain the observed star formation.

\subsection{What Went Wrong} \label{sect:WhatWhy}

For more than thirty years a few entrenched errors have slowed progress in understanding the dynamics of disk galaxies. Here I highlight the two areas which have caused the most confusion.

\subsubsection{Rotation Curves}
Rotation curves are not measured directly - they are \emph{fit} using a set of modelling assumptions. 
\cite{Beg89,Sof01,McGau01,deBlo02,Ior17,Oh18} trace the evolution of the methods to determine rotation curves from single long slit spectra measurements to much more detailed methods which include 2-D (and even 3-D) fits to detailed velocity-position maps in \Ha, HI, and CO.
For the standard 2-D tilted ring model, the galaxy center and velocity are held fixed and each ring element is determined separately, characterized by circular velocity and inclination.
The methodology makes the explicit simplifying assumption that the velocities form circular orbits, excluding the possibility of a non-symmetric or expanding disk from the outset.

Standard methods do not measure rotation but rather the azimuthal velocity of escaping material resulting in a mismatch between observed galaxy mass and the dynamical mass required by the erroneous `rotation' curves.

A few recent analyses \citep{Schm16,DiTeo21,Sel21} relax the assumption of perfectly circular orbits somewhat to search for radial flows.
First the \HI\ data is fit using the standard tilted ring model and then the radial flow component which minimizes error is found.
The basic problem with this technique is that it's possible to find a excellent fit to an expanding disk's velocity field with a tilted ring model.
If a good fit to the data is obtained in the first step, which is common, the search for radial flows fails because no radial flow can improve the fit. In fact, \cite{Sel21} analysed the \HI\ velocity data for NGC 3198 and found that an excellent fit without radial flows. The bar-driven flows apparent in Figures \ref{fig:3198Spitzer-5ColorBar} and   \ref{fig:3198fourfigs} were simply missed.

\cite{Syl19} discuss the failure of the tilted ring model in the presence of non-axisymmetric or radial flows
and gives results using a new galaxy template that allows for the case where circular velocity dominates the inner disk regions but radial velocities are dominant in the outskirts.
The new template was applied to the galaxy sample of 19 galaxies used in \cite{deBlo08} augmented with an additional 9 galaxies which appear to be dominated by radial flows in the outer regions.
The results were excellent.
The fit of the new template for NGC 3198, using fewer free variables, is nearly as good as the tilted ring mode but, of course, describes a completely different velocity field.

The tilted ring model has been used for the outer core with insufficient justification. Historically, two sorts of verification were used to verify the model. First, it was verified that velocity curve based on \HI\ data matched the velocity curves from CO or \Ha\ in the inner disk. This was taken to mean that the extended \HI\ velocity curve was correct. The second sort of verification generated sample \HI\ data based on an assumed axisymmetric rotating disk and showed that the tilted ring model could reproduce the assumed data quite accurately. The problems with these these approaches are obvious in hindsight.

Barless late-type galaxies also lose gas outwardly but not as dramatically as barred galaxies. The result is that the stellar disk sits at the center of a large, expanding, gaseous disk. The azimuthal velocity of the \HI\ gas increases to the edge of the disk and is then nearly constant and it is impossible to determine the expansion rate of the disk only from the \HI\ velocities.

\subsubsection{Exponential Disks}
Late type disk galaxies exhibit exponential light distribution over the stellar disk and in addition face-on galaxies are exponential in the outer regions. This has been taken to mean that the mass distribution is exponential to great distances so that Freeman's solution \citep{Fre70} for an infinite flat exponential disk is a good approximation for the gravitational attraction of the stellar disk leading to the frequently used `maximum disk' methodology to fit rotation curves.

As discussed in Section \ref{sect:AreTruncated}, disk galaxies do not extend to infinity but rather are truncated at the edge of the stellar disk.  Any mass beyond the truncation radius is only a fraction of the stellar disk and is warped and flared so that this outer region has negligible effect on the gravitational field. The observed exponential fall of the stellar population in the outer region is due to conflating the disk population with the stellar halo which, being pressure supported, falls off exponentially.

The `maximum disk' methodology attempts to find the maximum gravitational contribution of the observed baryonic mass by using the exponential disk solution and a bounding mass-to-light ratio \citep{Bos81::1,Bos81::2,vanAlb86,Beg89,Beg91}. The methodology fails completely if the rotation curves used in the calculations don't actually measure rotation. Furthermore, it is often overlooked  that the gravitational attraction of an infinite exponential disk is a factor of 2-3 less than that of a truncated disk and in fact is less than that of a point mass of equal mass.

\section{Summary and Conclusions} \label{sect:Summary}
Disk galaxies grow by transporting radial momentum outwardly until reaching a condition of near instability. The velocity at the edge these galaxies is greater than for a point mass by a factor of 2 or more and the orbital velocity can approach the escape velocity of the system. These unstable galaxies are susceptible to losing mass outwardly by thermal escape or by gross outflow initiated by a bar or a close encounter with another galaxy.

The instability at the edge of disk galaxies can be seen in flat disk models which follow strictly Newtonian physics.
For these disks, the radial derivative of total energy of an orbiting mass is \emph{negative} in the region just beyond the edge of the disk.  That is, the object can gain energy by moving to a higher orbit and so no stable orbit exists in this region. Any mass found in the region is just passing through.

Spiral galaxies are truncated at a radius  somewhat beyond the Holmberg radius where the surface density drops below \app1 \Msun\ pc$^{-2}$. The truncation is most easily seen in edge-on spirals but is often obscured in face-on spirals by halo stars which fall off exponentially. The truncation is not complete and stars and gas form an outer disk consisting of the arms of grand design spirals or the symmetric expanding \HI\ disks of LSB galaxies. `Rotation' curves determined from velocity measurements of the outer \HI\ do not measure rotation as such but rather, these curves measure the azimuthal velocity of gas escaping the almost unimpeded by the gravity of the light inner disk.

A characteristic gravitational attraction, Milgrom's constant, is observed at the edge of spiral galaxies and is the threshold for gas loss. If the disk expands such that the radial attraction at the edge of the disk is less than Milgrom's constant, the rate of gas loss increases, shutting down star formation and stopping the expansion. This process explains the several strong correlations of mass, radius and velocity of disk galaxies.

The attraction at the edge of the stellar disk is greater than that of a spherical mass and so material in orbit at the edge of the disk moves faster and has greater kinetic energy than a particle in orbit around a spherical mass. The orbital velocity of a particle at the disk edge can approach or even exceed escape velocity and material which escapes from such a disk moves outward with nearly constant velocity. The radial component of the velocity is missed by standard fitting procedures.

For realistic truncated disk models, the radial derivative of the total energy of an orbiting particle is negative in the region just beyond the disk edge and stable orbits are not possible in this region.  This is very different from the behaviour of the total energy for a particle in the field of a spherical mass which increases monotonically with radius so that energy must be supplied to reach a higher orbit. This instability near the disk edge explains the sharp truncations observed for late-type disks seen edge-on.

For decades, a few basic misunderstandings have affected the methodology used to study galaxy dynamics. Progress has stalled because the errors are self-reinforcing and are now deeply imbedded in the literature. Despite the many problems that affect the CDM model it remains standard paradigm because the only alternatives on offer are incomplete or non-physical expansions of the fundamental laws of physics.

The need for dark matter to explain galaxy rotation curves arises because of the reasonable but mistaken assumption that the movement of stars and gas are governed in the same way in the outer disk region as in the inner regions. However, I have shown here that this is not the case; the stars and gas in the outer parts of galaxies which move much faster than expected are simply not in stable circular orbits. Newtonian physics, properly applied, are sufficient to resolve the issues.

\section*{Acknowledgements}

This work extends the investigations of Sylos Labini, Benhaiem and their collaborators cited above who take a different approach to the problems. 

This research used a Spitzer Space Telescope Infrared Array Camera (IRAC) 3.6 micron Super Mosaic image from the  set of Enhanced Imaging Products (SEIP) from the Spitzer Heritage Archive.

This research has made use of the Aladin interactive sky atlas, developed at CDS, Strasbourg, France.

This research made use of public domain program AstroImageJ developed Karen Collins and John Kielkopf, University of Louisville based on ImageJ developed by Wayne Rasband National Institute of Health.  %from program help page

% From  https://classic.sdss.org/dr6/coverage/credits.html:
This work made use of the Sloan Digital Sky Survey (SDSS).
    Funding for the SDSS and SDSS-II has been provided by the Alfred P. Sloan Foundation, the Participating Institutions, the National Science Foundation, the U.S. Department of Energy, the National Aeronautics and Space Administration, the Japanese Monbukagakusho, and the Max Planck Society, and the Higher Education Funding Council for England.

This work made use of products of the Galaxy Evolution Explorer(Galex) mission.
    Caltech leads the Galaxy Evolution Explorer mission and is responsible for science operations and data analysis. NASA's Jet Propulsion Laboratory, Pasadena, Calif., manages the mission and built the science instrument. The mission was developed under NASA's Explorers Program managed by the Goddard Space Flight Center, Greenbelt, Md. South Korea and France are the international partners in the mission.

\section*{Data Availability}
Figure \ref{fig:3198Spitzer-5ColorBar}    uses  3.6 \mm\ Super Mosaic image from the Spitzer Enhanced Image Program(SEIP).
File 50062440.50062440-0.IRAC.1.mosaic.fits is archived at DOI:\href{https://doi.org/10.26131/IRSA433}{10.26131/IRSA433}.

 Figure 6.a shows NGC 3198 in SDSS colourized optical.
The original SDSS9  u,g,r,i and z band FITS files are archived on the \href{https://dr12.sdss.org/}{DR12 Science Archive Server} and can be accessed at the \href{https://dr12.sdss.org/fields/name?name=ngc+3198}{DR12 Science Archive Server.}

Figure 6.b is a Galex colourized ultraviolet image which was produced from data archived at MAST DOI: \href{https://doi.org/10.17909/hkn3-fv52}{10.17909/hkn3-fv52}.

Figures 6.c and  6.b show VLA data generated for the THINGS project \citep{Ken03} and archived at the \href{https://www2.mpia-hd.mpg.de/THINGS/Overview.html}{THINGS Data Repository}.

\section*{Code availability}
No new computer codes were used in this work.

The line plots of Figures 1, 2, 3, 4 and 7 were generated using Maple2023 and the Maple worksheets, which include the defining equations for the several flat disks discussed here, are available \href{https://drive.google.com/drive/folders/1mWrCh3H2bCDc3O_eYQGeUs17YMSQ1oLX?usp=drive_link}{here}.

\section*{Competing interests}
The author declares no competing interests.

\appendix \label{AppendixA}

\section[\appendixname~\thesection]{The force and potential of a cylinder}\label{secA1}
%\subsection[\appendixname~\thesubsection]{}

The goal  is to find the force and potential along the axis of a homogeneous right cylinder. The solution for the force is well known
but the solution for the potential is not.
To begin, consider the force due to a thin ring of mass $M_r$ and radius $R_r$ on the y-z plane centered on x=0. The force at a point  the x axis can be written immediately:
\begin{equation}
  F_{ring} = -\frac{M_{ring} G x}{ (x^2+R_{ring}^2)^{3/2}}
\end{equation}

Use this result as a kernel to find the force due to thin disk in the y-z plane centered on x=0.
\begin{align}
  F_{disk} &= \int^{R_{disk}}_0 { - \frac{2 \pi r \sigma G x}{(x^2+r^2)^{3/2}}}dr  \nonumber    \\
           &=  -\frac{2 M_{disk}  G( \sqrt{x^2 + R_{disk}^2  } -x )   }        { R_{disk}^2 \sqrt{x^2 + R_{disk}^2  }}
\end{align}
where the mass of the disk is $M_{disk} = \pi \sigma R_{disk}^2 $.
Now find the attraction of a solid bar with total length $\lambda$, density $\rho$, radius $\alpha$, positioned between x=0 and x=${\lambda}$  at a point on the axis a distance h beyond the end of the bar:
\begin{align}
 F_{half}(h,{{\lambda}}) &=  \int^{{\lambda}}_0 -\frac{2 \pi \rho G( \sqrt{\alpha^2 +(\widehat{x}+h)^2} -\widehat{x} -h )}{\sqrt{\alpha^2 +(\widehat{x}+h)^2} }d\widehat{x}   \nonumber    \\
          &=  -2 \pi \rho G ( \sqrt{\alpha^2+h^2} -\sqrt{\alpha^2+(h+{{\lambda}})^2} + {{\lambda}} )
\end{align}
The force of a homogeneous bar with radius $\alpha$, length 2L and density $ \rho =  {M_{bar}}/({ 2 \pi \alpha^2 L})$  positioned between \hbox{-L\le x\le L} is found by using the expression for $F_{half}$.
\begin{align}
F_{barO}(x) &= F_{half}(x-L,2L)                                                               \nonumber\\
           &=  - \frac{M_{bar} G}{\alpha^2L}  ( \sqrt{ (L-x)^2 +\alpha^2}  -\sqrt{ (L+x)^2 +\alpha^2}  +2L )       &  L< x  \\
F_{barI}(x) &= F_{half}(0,L+x) - F_{half}(0,L-x)                                            \nonumber \\
        &=  - \frac{M_{bar} G}{\alpha^2L}  ( \sqrt{ (L-x)^2 +\alpha^2}  -\sqrt{ (L+x)^2 +\alpha^2}  +2x )       &  ~~ 0< x < L
\end{align}

The potential is found by integrating the force to infinity. For points beyond the bar this is:
\begin{align}
\Phi_{barO}(x)        = & \int^\infty_x  F_{barO}  (\widehat{x})  d\widehat{x}    \nonumber\\
\Phi_{barO}(x)        = & - \frac{M_{bar} G}{2 \alpha^2L}  \biggl[   \alpha^2\bigl( \arcsinh\bigl(\frac{x+L}{\alpha}\bigr)-\arcsinh\bigl(\frac{x-L}{\alpha}\bigr)\bigr)    \nonumber\\
                        & +(x+L)\sqrt{(x+L)^2 +\alpha^2}  \nonumber\\
                        & -(x-L)\sqrt{(x-L)^2+\alpha^2}  -4xL   \biggr]   ~~~~~& L< x
\end{align}
The potential for points within the bar is:
\begin{align}
\Phi_{barI}(x)        = & \int^L_x  F_{barI}  (\widehat{x})  d\widehat{x}    +  \Phi_{barO}(L)   \nonumber\\
\Phi_{barI}(x)        = & - \frac{M_{bar} G}{2 \alpha^2L}  \biggl[   \alpha^2\bigl( \arcsinh\bigl(\frac{L+x}{\alpha}\bigr)  +\arcsinh\bigl(\frac{L-x}{\alpha}\bigr)\bigr)   \nonumber\\
                        & +(L+x)\sqrt{(L+x)^2 +\alpha^2}  \nonumber\\
                        & +(L-x)\sqrt{(L-x)^2+\alpha^2}  -2L^2 -2x^2   \biggr]           & 0<x<L
 \end{align} $\Box$

%\bibliography{ejs-Outflowbib}

\bibliographystyle{plainnat}

\end{document}